\documentstyle[twocolumn,prl,epsf,aps]{revtex}

\newcommand{\rmd}{{\rm d}}
\newcommand{\rmi}{{\rm i}}

\newcommand{\beq}{\begin{equation}}
\newcommand{\eeq}{\end{equation}}
\newcommand{\bea}{\begin{eqnarray}}
\newcommand{\eea}{\end{eqnarray}}

\newcommand{\lsr}{\langle\sigma\rangle}
\newcommand{\llr}{\langle\lambda\rangle}

\newcommand{\lanthan}{La$_{2}$CuO$_{4}$ }
\newcommand{\ybco}{YBa$_{2}$Cu$_{3}$O$_{6.15}$ }

\begin{document}                                                

\wideabs{

\draft

\title{Effective Sublattice Magnetization and N\'eel Temperature
in Quantum Antiferromagnets}

\author{Eduardo C. Marino and Marcello B. Silva Neto}

\bigskip

\address{
Instituto de F\'isica, Universidade Federal do Rio de Janeiro,\\
Caixa Postal 68528, Rio de Janeiro - RJ, 21945-970, Brazil
}

\date{\today}
\maketitle


\begin{abstract} 

We present an analytic expression for the finite temperature effective 
sublattice magnetization which would be detected by inelastic neutron 
scattering experiments performed on a two-dimensional square-lattice
quantum Heisenberg antiferromagnet with short range N\'eel order. Our 
expression, which has no adjustable parameters, is able to reproduce both 
the qualitative behaviour of the phase diagram $M(T)\times T$ and the 
experimental values of the N\'eel temperature $T_{N}$ for either doped 
YBa$_{2}$Cu$_{3}$O$_{6.15}$ and stoichiometric La$_{2}$CuO$_{4}$ compounds. 
Finally, we remark that by incorporating frustration and $3D$ effects 
as perturbations is sufficient to explain the deviation of the experimental 
data from our theoretical curves.

\end{abstract}


}

\begin{narrowtext}


Two dimensional quantum antiferromagnetism has been a matter of great 
interest and subject to intense investigation, due to its possible relation 
to the normal state properties of high-temperature superconductors. There is 
by now clear experimental evidence that the pure high-$T_{c}$ superconducting 
cuprate compounds are well described by a quasi two-dimensional $S=1/2$ 
Heisenberg antiferromagnet on a quasi-square lattice, whose sites are occupied 
by $Cu^{++}$ magnetic ions. The dynamical structure factor of the $2D$ Heisenberg 
antiferromagnet, calculated via the mapping of the Heisenberg Hamiltonian onto 
the $O(3)$ nonlinear sigma model \cite{Chakravarty2}, was successfuly confirmed 
by inelastic neutron scattering experiments on La$_{2}$CuO$_{4}$ \cite{Yamada}. 
Several other microscopic techniques like light scattering \cite{Light-Scat}, 
muon spin relaxation \cite{Muon} and thermal neutron scattering 
\cite{Thermal-Neutron}, have also been used to probe the magnetic 
correlations in these materials and confirmed the quasi $2D$ Heisenberg 
antiferromagnet hypotesis.

A common feature among almost all superconducting cuprate compounds is the 
existence of a N\'eel ordered moment in the low temperature, underdoped regime. 
As the temperature is increased, or the sample doped, antiferromagnetic order 
is destroyed, leading to new forms of spin order \cite{Marino}. According to 
spin-wave theory, for a $\rmd$-dimensional hypercubic lattice, N\'eel order is 
possible at $T=0$ for $\rmd\geq 2$. However, despite the widespread success of 
spin-wave theory, there remain a number of issues that defy the description of 
the superconducting cuprate compounds within this approach. For example, the 
development of a sublattice magnetization is known to be suppressed in the 
two-dimensional Heisenberg antiferromagnet for any nonzero temperature. True 
long range order, as a genuine three-dimensional phenomenon, would only be 
achieved by considering the interlayer coupling 
$J_{\perp}\sim 10^{-5} J_{\parallel}$ not only as a perturbation.

It is the purpose of this work to show that the experimental data for the sublattice 
magnetization of \lanthan \cite{PRB-Keimer} and \ybco \cite{Tranquada} can in fact 
be described still in the context of a two-dimensional square-lattice quantum 
Heisenberg antiferromagnet at finite temperatures, as far as inelastic neutron 
scattering experiments are concerned. Our starting point is the observation that
the nature of the spin correlations in the renormalized classical regime is consistent 
with one of the three possibilities of fig. \ref{Fig-Ren-Class}, according to the observation 
wave vector $|k|$, or frequency $\omega$ \cite{Sachdev}. In this sense, 
any possible neutron scattering experiment, with high enough energy transfers, 
performed on a true two-dimensional system, would actually measure a nonvanishing 
effective sublattice magnetization, since one would be probing the dynamics of spin 
correlations in the intermediate Goldstone region. Inelastic neutron scattering 
experiments probe a microscopic, short wavelength physics to which we can associate an 
effective N\'eel moment. As it will become clear, the behaviour of this effective moment 
can be described by an effective field theory for the low frequecy, long wavelength fluctuations
of the spin fields about a state with short range N\'eel order. We will then be able to speak about 
a finite temperature phase transition in the $2D$ system, associated to the colapse 
of the Goldstone region in fig. \ref{Fig-Ren-Class}.

%
\begin{figure}[h]
\centerline{\epsfxsize=7cm \epsffile{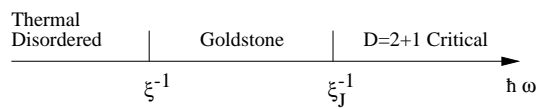}}
\caption{Properties of the $2D$ quantum Heisenberg antiferromagnet as a 
function of the frequency $\omega$. $\xi$ is the actual correlation length while 
$\xi_{J}$ is the Josephson correlation length related to the spin-stiffness
by $\xi_{J}=\hbar c/\rho_{s}$.}
\label{Fig-Ren-Class}
\end{figure}
%

The two-dimensional square-lattice quantum Heisenberg antiferromagnet has a well
known continuum limit given in terms of the $2+1$ dimensional $O(3)$ quantum 
nonlinear sigma model \cite{Haldane}. The later, on the other hand, is defined by 
the partition function
\beq
{\cal Z}(\beta)=\int{\cal D}n_{l}{\;}\delta(n_{l}^{2}-1){\;}
\exp{(-{\cal I}(n_{l}))},
\label{Part-Func}
\eeq
where the action
\beq
{\cal I}(n_{l})=\frac{\rho_{0}}{2\hbar}
\int_{0}^{\beta\hbar}\rmd\tau\int\rmd^{2}{\bf x}
\left[(\nabla n_{l})^{2}+
\frac{1}{c_{0}^{2}}(\partial_{\tau} n_{l})^{2}
\right]
\label{Action}
\eeq
describes the long-wavelength fluctuations of the staggered components of the 
spin-field $n_{l}=(\sigma,\vec{\pi})$, $l=1,\dots,N=3$. The fixed length 
constraint is understood. In the above expression, $\rho_{0}$ is the spin stiffness, 
$c_{0}$ is the spin-wave velocity, $\beta=(k_{B}T)^{-1}$ and all quantities with a 
$0$ subscript represent bare quantities.

We shall work in the natural units $k_{B}=\hbar=c=1$, with $c$ being the 
renormalized spin wave velocity. \footnote{For large $N$ the spin wave velocity 
does not renormalize and $c_{0}=c$.} Also, further analysis will be simply 
expressed in terms of the coupling constant $g_{0}=N/\rho_{0}$, which 
has the units of inverse length.  With this notation and choosing the staggered 
magnetization to be along the $\sigma$ field direction, we can integrate over 
the remaining $N-1$ spin-wave degrees of freedom $\vec{\pi}$ and study the 
behavior of the partition function (\ref{Part-Func}) in the large $N$ limit. 
As usual, $N$ is taken to be large enough while $g_{0}$ is kept fixed. This 
means that we have to choose $\rho_{0}\sim N$.

For large $N$, the partition function (\ref{Part-Func}) is dominated by the 
stationary configurations of the magnetization, $\lsr$, and of the Lagrange 
multiplier field, $\rmi\llr=m^{2}$, introduced in order to ensure the averaged
fixed length constraint. These, on the other hand, can be determined from the 
stationarity conditions 
\bea
m^{2}\lsr&=&0, \nonumber \\
\lsr^{2}&=&\frac{1}{g_{0}}-\frac{1}{\beta}
\sum_{n=-\infty}^{\infty}\int_{0}^{\Lambda}\frac{\rmd^{2}{\bf k}}{(2\pi)^{2}}
\frac{1}{{\bf k}^{2}+\omega_{n}^{2}+m^{2}},
\label{Saddle-Point-Eq}
\eea
where a cutoff $\Lambda$ was introduced to make the momentum integral 
ultraviolet finite.

From the above set of equations we see that the antiferromagnetic system
could in principle be found in two distinct phases. If $\lsr\neq 0$ 
then $m=0$ and the system would be in the Goldstone phase with a 
nonvanishing net sublattice magnetization. In this case the ground state 
would exhibit true long range N\'eel order. If $m\neq 0$ on the other hand, 
then $\lsr=0$ and N\'eel order is absent. There are no 
gapless excitations in the spectrum of the finite temperature system. It 
is a well known fact that for the $2+1$ dimensional $O(N)$ invariant 
nonlinear $\sigma$ model at $T>0$, the only possible physical situation is 
the second one, due to severe infrared divergencies in the second saddle-point 
equation (\ref{Saddle-Point-Eq}). As a consequence, the value of $m$ is pushed 
from zero to a finite value making the sublattice magnetization $\lsr$ to 
vanish, in agreement with the Coleman-Mermin-Wagner theorem.

We can compute the value of the $O(N)$ invariant mass $m$ in a closed form 
by subtracting the linear divergence in the second saddle-point equation 
(\ref{Saddle-Point-Eq}) as 
\beq
\frac{1}{g_{0}}=\frac{1}{g_{c}}+\frac{\rho_{s}}{4\pi N},
\label{Renormalization}
\eeq
where $g_{c}=4\pi/\Lambda$ is the bulk critical coupling and 
\beq
\rho_{s}=\frac{\sqrt{S(S+1)}}{2\sqrt{2}}\frac{\hbar c}{a},
\label{Rho-from-c}
\eeq
with $S=1/2$ and $a$ being the lattice spacing \cite{Haldane}.
Now, after momentum integration and frequency sum we arrive at 
\beq
\xi^{-1}=m(\beta)=\frac{2}{\beta}{\rm arcsinh}{\left(\frac{e^{-\beta\rho_{s}/(2N)}}{2}\right)},
\label{Dinamical-Mass}
\eeq
which is nonvanishing for $T>0$, thus indicating that a N\'eel phase can 
only occur at $T=0$. The conclusion is that even at the smallest temperature 
there is a gap in the spin-wave spectrum and a finite correlation length 
which measures the size of clusters in which there is short range N\'eel 
order. The above expression for $\xi$ has been obtained by Chakravarty
{\it et al.} \cite{Chakravarty2} and its zero temperature limit successfully 
confirmed by quasi-elastic neutron scattering experiments on \lanthan \cite{Endoh}, 
for, however, temperatures approaching $T_{N}$ from above, $T\rightarrow T_{N}^{+}$, 
where this compound is known to exhibit a true two-dimensional behaviour. 

Let us now consider inelastic neutron scattering experiments, performed on a true 
$2D$ system, with energy transfers $\Delta E=\hbar\omega$ such that the corresponding 
wavelength, $\lambda=1/\hbar\omega$, satisfies $\xi_{J}\leq\lambda\leq\xi$. Typical time 
scales in such experiments are $\tau_{\lambda}=\lambda/c$, consequently much smaller 
than the relaxation time $\tau=\xi/c$ at which the $2D$ system disorders. For such 
experiments, spins would look like as if they were frozen and a nonvanishing effective 
sublattice magnetization would be measured. Since at low temperatures $\xi$ is much larger 
than $\xi_{J}$, the three regions of fig. \ref{Fig-Ren-Class} are well separated. In the 
large intermediate region, probed by our experiment, the system behaves as if it had true 
long range antiferromagnetic order and dynamic scaling hypotesis is justified \cite{Chakravarty2}.  
We are then allowed to apply a hydrodynamic picture for the low frequency, long-wavelength 
fluctuations of the spin-fields $n_{l}$, in which its short-wavelength fluctuations follow 
adiabatically the fluctuations of the disordered background whose typical wavelenth is the 
scale of disorder, the correlation length. The effective field theory to describe the spin 
correlations in this intermediate Goldstone region is obtained by functionally integrating 
the Fourier components of the fields in (\ref{Part-Func}) with frequency inside momentum shells 
$\kappa\leq |\vec{k}| \leq\Lambda$. The resulting partition function is such that, for large $N$, 
the leading contribution now comes from the scale dependent stationary configurations $\lsr_{\kappa}$ 
and $\rmi\llr_{\kappa}=m_{\kappa}^{2}$, solutions of the new set of saddle-point equations
\bea
m_{\kappa}^{2}\lsr_{\kappa}&=&0, \nonumber \\
\lsr_{\kappa}^{2}&=&
\frac{1}{g_{0}}-\frac{1}{\beta}
\sum_{n=-\infty}^{\infty}\int_{\kappa}^{\Lambda}
\frac{\rmd^{2}{\bf k}}{(2\pi)^{2}}
\frac{1}{{\bf k}^{2}+\omega_{n}^{2}+m_{\kappa}^{2}}.
\label{New-Saddle-Point-Eq}
\eea
Differently from the previous case, now we can in fact find the system in
two different phases (regimes): ordered (asymptotically free) or disordered
(strongly coupled); depending on the size of $\xi_{\kappa}=1/\kappa$ relative 
to $\xi$: smaller (high energies) or larger (low energies). In the ordered 
phase, $\xi_{\kappa}\ll\xi$, $m_{\kappa}=0$ is the solution that minimizes the
free energy and the $2D$ system is then characterized by a nonvanishing 
effective sublattice magnetization $\lsr_{\kappa}\neq 0$, a divergent
effective correlation length $\xi_{eff}=1/m_{\kappa}=\infty$ and gapless
excitations in the spectrum. 

The effective sublattice magnetization can be exactly computed from the
second saddle point equation (\ref{New-Saddle-Point-Eq}). Using the
renormalization scheeme defined by (\ref{Renormalization}), we obtain, 
after momentum integration and frequency sum, the expression
\beq
\frac{\rho_{s}(\kappa,\beta)}{2\pi N}\equiv
\lsr_{\kappa}^{2}=\frac{\rho_{s}}{4\pi N}+
\frac{1}{2\pi\beta}\ln{(2\sinh{(\beta\kappa/2)})},
\label{Running-rho}
\eeq
which depends on the energy scale $\kappa$ and on the temperature.

Some comments are in order. The running spin stiffness (\ref{Running-rho}) decreases 
($g(\kappa,\beta)=N/\rho_{s}(\kappa,\beta)$ increases) as $\xi_{\kappa}\rightarrow\xi$, 
for a given temperature. This is a consequence of the fact that we are coarsing over 
degrees of freedom which actually feels the finite size of the clusters with short range 
N\'eel order. Furthermore, for $\xi_{\kappa}>\xi$ we would be coarsing over degrees of 
freedom outside these clusters, leading to the disordered (strongly coupled) phase. 
Lowering the scale $\kappa$ is also equivalent to waiting longer for a response, and for 
$\tau_{\kappa}\geq\tau$ we would be waiting long enough for the system to disorder. Here, 
conversely, in order to obtain a finite temperature phase transition in the $2D$ system, 
we will rather fix the scale $\kappa$ and study the behaviour of the effective spin 
stiffness (\ref{Running-rho}) with the running parameter being the temperature. We must 
fine tune, and hold fixed, the energy transfers in our experiment so that our $2D$ system 
is able to reproduce the observed $3D$ behaviour in real materials. For this it suffices 
to impose the boundary condition 
\beq
\rho_{s}(\kappa,T=0)=\rho_{s},
\label{Bound-Cond}
\eeq  
with $\rho_{s}$ being the bulk spin stiffness of the real system. From (\ref{Bound-Cond}) 
we conclude that $\kappa=\rho_{s}/N$, which is exactly the inverse Josephson correlation
length, $\kappa=\xi_{J}^{-1}$. This should not be surprising since the spin stiffness
is itself a microscopic, short wavelength quantity defined at the Josephson scale. Notice
also that $\kappa=23$ meV for the case of \lanthan, which is actually consistent 
with the energy transfers commonly used in this kind of experiment \cite{Yamada}. Now, 
inserting (\ref{Bound-Cond}) in (\ref{Running-rho}), the expression for the finite temperature 
effective sublattice magnetization, $M(T)\equiv\rho_{s}(\kappa=\rho_{s}/N,T)$, 
becomes 
\beq
M(T)=\frac{M_{0}}{2}+NT
\ln{\left(2\sinh{\left(\frac{M_{0}}{2NT}\right)}\right)},
\label{Magnetization}
\eeq
with $M_{0}=\rho_{s}$. The sublattice magnetization $M(T)$ vanishes at a N\'eel temperature 
$T_{N}$ given by
\beq
T_{N}=\frac{M_{0}}{N\ln{2}}.
\label{Critical-Temperature}
\eeq
For temperatures above $T_{N}$, we would be coarsing over degrees of freedom outside the 
shrinked clusters of size $\xi$, leading again to the disordered (strongly coupled) phase.

Let us now show that the above analysis can in fact be used to describe the experimental 
data for different cuprate compounds. Take for example the data obtained for \lanthan
\cite{PRB-Keimer} and \ybco \cite{Tranquada}. For these compounds, we find agreement 
between our predictions and the observed N\'eel temperatures, within almost $10 \%$, 
already at the leading order, as can be seen from table \ref{Table}. More important, 
we have obtained a good qualitative agreement between the phase diagram $M(T)\times T$ 
and experiment, for the whole range of temperatures from $0$ to $T_{N}$ 
(see dotted lines in figs. \ref{Fig-LaCO} and \ref{Fig-YBCO}). 

In order to have a flavor on how our results can be improved, let us mention that already at 
the next-to-leading order in the $1/N$ expansion we will have a nontrivial renormalization of 
the spin-wave velocity due to the self interaction between spin-waves \cite{Sachdev}. 
This should lower the value of $c$ and, if we take for example a lowering of about $10 \%$, we 
obtain the behaviour described by the solid curves in figs. \ref{Fig-LaCO} and \ref{Fig-YBCO}. 
The spin wave velocity will also be renormalized by dynamic scaling, but we assume that at
the shortest distances, that is $\hbar\omega\gg\xi^{-1}$, this effect can be neglected when 
compared to the effects of the self interactions. For this reason, it should not lead to a 
further damping of the spin waves.

%
\begin{table}[h]
\begin{tabular}{c|c|c|c|c}
${\;}$ & $c$ (eV $\AA/\hbar$) & $\rho_{s}$ (meV) & $T_{N}$ (K) & $T_{N}^{exp}$ (K)\\
\hline
YBa$_{2}$Cu$_{3}$O$_{6.15}$ & $1.00\pm 0.05$ & $81\pm 4$ & $450\pm 20$ & $410\pm 3$\\
${\;}$			    & $0.90\pm 0.05$ & $73\pm 4$ & $410\pm 20$ & ${\;}$ \\
\hline
La$_{2}$CuO$_{4}$ & $0.85\pm 0.03$ & $68\pm 2 $ & $380\pm 10$ & $325\pm 5$ \\
${\;}$		  & $0.75\pm 0.03$ & $60\pm 2 $ & $330\pm 10$ & ${\;}$ \\
\end{tabular}
\caption{To compute $\rho_{s}$ we have made used of formula (\ref{Rho-from-c}) 
with $a\sim 3.8 $ $\AA$ and $S=1/2$, while to compute $T_{N}$ we have used formula 
(\ref{Critical-Temperature}) with $N=3$. The experimental values of $c$ were taken
from ref. \protect\cite{Kampf}.}
\label{Table}
\end{table}

Notice now that, with respect to the solid curves, the experimental points can be separated 
into two different sets. For $T<T_{N}/2$, we find all points below the solid curves while, 
for $T>T_{N}/2$, we find, instead, the points all above our theoretical prediction. This is 
consistent with a picture in which strong frustration induced quantum fluctuations, due for 
example to a nonzero next-nearest-neighbour coupling, are dominant at low temperatures and 
suppressed at higher $T$, where the effects due to a sizable $J_{\perp}$ begin to be felt. 
Notice also that in the case of \ybco the points deviate even more from the solid curve, for 
$T>T_{N}/2$, than in the case of \lanthan. We attribute this to the bilayer structure of \ybco, 
which causes a further increase in the sublattice magnetization. As we approach $T_{N}$ from 
below, both systems behave effectively as true $2D$ Heisenberg antiferromagnets with 
nearest-neighbour coupling, as shown by the experimental data. From the above discussion we 
conclude that by incorporating frustration and $3D$ effects as perturbations, with properly 
temperature renormalized coeficients, might be sufficient to account for the deviation of the 
data from our theoretical predictions. We are presently investigating this possibility.

%
\begin{figure}[h]
\centerline{\epsfxsize=6cm \epsffile{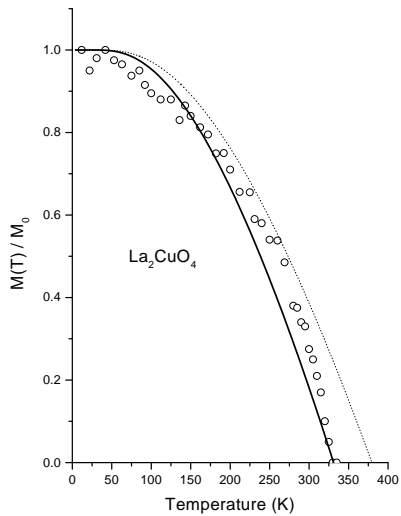}}
\caption{For the dotted line we have used $\hbar c=0.85$ eV $\AA$ while for the 
solid line we assumed $\hbar c = 0.75$ eV $\AA$. Experimental data from ref. 
\protect\cite{PRB-Keimer}.}
\label{Fig-LaCO}
\end{figure}
%

As a final remark, let us show that our treatment is consistent with experiment also 
in the disordered phase. Above $T_{N}$, where the effective sublattice magnetization 
vanishes, we must consider the second possible solution for the set of saddle-point 
equations (\ref{New-Saddle-Point-Eq}), namely $\lsr_{\kappa}=0$ and $m_{\kappa}\neq 0$. 
If we then solve the self-consistent equation for the gap we end up with
\beq
m^{2}_{\kappa}=\frac{4}{\beta^{2}} {\rm arcsinh}^{2}
{\left(\frac{e^{-\beta\rho_{s}/(2N)}}{2}\right)}-\left(\frac{\rho_{s}}{N}\right)^{2},
\label{Effective-Gap}
\eeq
for $\kappa=\rho_{s}/N$. It is straightforward to see that the effective correlation 
length $\xi_{eff}=1/m_{\kappa}$ diverges as $T\rightarrow \rho_{s}/(N\ln{2})$, or
in other words, as we approach $T_{N}$ from above. This is consistent with the data for
\lanthan from \cite{Endoh}, and for the correct temperature limit.

%
\begin{figure}[h]
\centerline{\epsfxsize=6cm \epsffile{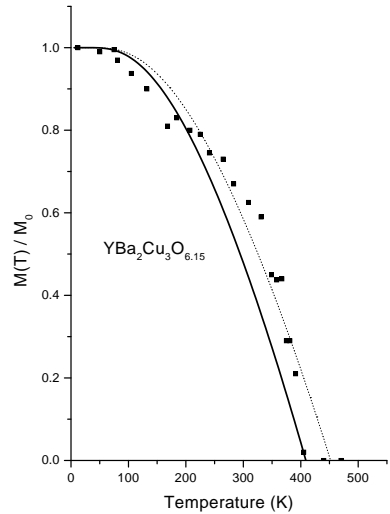}}
\caption{For the dotted line we have used $\hbar c=1.00$ eV $\AA$ while for the 
solid line we assumed $\hbar c = 0.90$ eV $\AA$. Experimental data from ref. 
\protect\cite{Tranquada}.}
\label{Fig-YBCO}
\end{figure}


The authors have benefited from several fruitful discussions with C. Farina, 
A. Katanin, B. Keimer, E. Miranda and F. Nogueira. E.C.M. is partially supported 
by CNPq and FAPERJ. M.B.S.N is supported by FAPERJ.

\end{narrowtext}



\begin{references}

\bibitem{Chakravarty2}
S. Chakravarty {\it et al.}, Phys. Rev. {\bf B39},
2344 (1989).

\bibitem{Yamada}
K. Yamada {\it et al.}, Phys. Rev. {\bf B40}, 4557 (1989).

\bibitem{Light-Scat}
K. Lyons {\it et al.}, Phys. Rev. {\bf B37}, 2353 (1988).

\bibitem{Muon}
D. Harshman {\it et al.}, Phys. Rev. {\bf B38}, 852 (1988).

\bibitem{Thermal-Neutron}
D. Vaknin {\it et al.}, Phys. Rev. Lett. {\bf 58}, 2802 (1987).

\bibitem{Marino} E. C. Marino, Phys. Lett. {\bf A263}, 446 (1999).

\bibitem{PRB-Keimer}
B. Keimer {\it et al.}, Phys. Rev. {\bf B45}, 7430 (1992).

\bibitem{Tranquada} J. M. Tranquada {\it et al.}, Phys. Rev. Lett. {\bf 60},
156 (1988).

\bibitem{Sachdev}
A. V. Chubukov {\it et al.}, Phys. Rev. {\bf B49}, 11919 (1994).

\bibitem{Haldane} F. D. M. Haldane, Phys. Rev. Lett. {\bf 50}, 1153 (1983).

\bibitem{Endoh} Y. Endoh {\it et al.}, Phys. Rev. {\bf B37}, 7443 (1988).

\bibitem{Kampf} 
A. P. Kampf, Phys. Rep. {\bf 249}, 219 (1994).

\end{references}
\end{document}